\documentclass[twocolumn,prb,showpacs,superscriptaddress,showkeys,preprintnumbers,amsmath,amssymb]{revtex4}

\usepackage{graphicx}
\usepackage{dcolumn}
\usepackage{bm}
\usepackage{multirow}
\usepackage{subscript}
\usepackage{nicefrac}

\begin{document}


\title{Anisotropy of incommensurate magnetic excitations in
slightly overdoped Ba$_{0.5}$K$_{0.5}$Fe$_2$As$_2$ probed by polarized
inelastic neutron scattering experiments}

\author{N. Qureshi}
\email[Corresponding author. Electronic
address:~]{qureshi@ph2.uni-koeln.de} \affiliation{$II$.
Physikalisches Institut, Universit\"{a}t zu K\"{o}ln,
Z\"{u}lpicher Strasse 77, D-50937 K\"{o}ln, Germany}

\author{C. H. Lee}
\author{K. Kihou}
\affiliation{National Institute of Advanced Science and Technology
(AIST), Tsukuba, Ibaraki 305-8568, Japan}

\author{K. Schmalzl}
\affiliation{J\"ulich Centre for Neutron Science JCNS, Forschungszentrum J\"ulich GmbH, Outstation at ILL, BP156, 38042 Grenoble, France)}

\author{P. Steffens}
\affiliation{Institut Laue Langevin, BP156X, 38042 Grenoble Cedex,
France}

\author{M. Braden}\email{braden@ph2.uni-koeln.de}
\affiliation{$II$. Physikalisches Institut, Universit\"{a}t zu
K\"{o}ln, Z\"{u}lpicher Strasse 77, D-50937 K\"{o}ln, Germany}

\date{\today}

\begin{abstract}

Polarized neutron scattering experiments on the slightly overdoped
superconductor Ba$_{0.5}$K$_{0.5}$Fe$_2$As$_2$ reveal broad
magnetic resonance scattering peaking at approximately 15 meV. In
spite of doping far beyond the suppression of magnetic order, this
compound exhibits dispersive and anisotropic magnetic excitations.
At energies below the resonance maximum, magnetic correlations
polarized parallel to the layers but perpendicular to the
propagation vector are reduced by a factor two compared to those
in the two orthogonal directions; in contrast correlations at the
peak maximum are isotropic.

\end{abstract}


\maketitle


The observation of enhanced magnetic fluctuations in the
superconducting phases of FeAs-based materials \cite{kam2008}
yields the strongest support for a pairing mechanism associated
with magnetism.\cite{chr2008} Collective resonance excitations
appearing just at the propagation vector of the antiferromagnetic
order in the parent phases were reported in polycristalline
optimally hole-doped Ba$_{0.6}$K$_{0.4}$Fe$_2$As$_2$
(Ref.~\onlinecite{chr2008}) as well as in  the electron-doped
BaFe$_2$As$_2$ series.\cite{lum2009,chi2009} Various neutron
scattering studies report the effect of electron doping on the
resonance peak by substituting Fe by either Co
(Refs.~\onlinecite{lum2009,par2009,ino2010,pra2010,par2010,li2010})
or Ni (Refs.~\onlinecite{chi2009,li2009,har2009,lip2010,wan2011}).
The energy of this excitation has been found to scale with $T_c$
and the intensity behaves like the superconducting ordering
parameter\cite{chi2009} while the dispersion of the resonance
seems to reflect the vicinity of the antiferromagnetic phase.
\cite{lee2013} A rather strong resonance has been detected in
single-crystalline optimally hole-doped
Ba$_{0.67}$K$_{0.33}$Fe$_2$As$_2$  ($T_c$=38 K).
\cite{wang2013,zhang2013}

Using polarized neutron scattering one may distinguish the
polarization direction of the fluctuating magnetic moments and
thereby directly detect possible spin-space anisotropies of
magnetic excitations. Such experiments were performed on Co
(Ref.~\onlinecite{ste2013}), on Ni-doped BaFe$_2$As$_2$
(Ref.~\onlinecite{luo2013}) and on
Ba$_{0.67}$K$_{0.33}$Fe$_2$As$_2$ (Ref.~\onlinecite{zha2013})
revealing clear evidence for strong spin-space anisotropy with a
universal scheme. Scanning the energy dependence of the magnetic
scattering at the fixed scattering vector ${\bf Q}$=(0.5 0.5
$q_l$) yields an isotropic signal just at the maximum of the total
scattering. This broad maximum is usually detected in unpolarized
neutron scattering experiments and labelled as the resonance mode.
At the lower energy side however, magnetic excitations are
anisotropic. Anisotropic scattering either appears in form of a
sharp isolated peak \cite{ste2013} or just as a shoulder
\cite{luo2013,zha2013} of the broad resonance feature. The three
directions relevant for the discussion of the spin-space
anisotropy in doped BaFe$_2$As$_4$ can be labelled as longitudinal
in-plane, i.e. parallel to the in-plane component of the
scattering vector, which we always chose as ${\bf Q}$=(0.5 0.5
$q_l$), [110], transverse in-plane, i.e. perpendicular to the
scattering vector and parallel to the planes [1$\bar{1}$0], and
out of plane. All studies \cite{ste2013,luo2013,zha2013} indicate
that the transversal in-plane polarized magnetic excitations only
appear at higher energies in the superconducting samples and that
they do not contribute to the anisotropic signal. This anisotropy,
therefore, closely resembles the observation of pure and
antiferromagnetic ordered BaFe$_2$As$_2$
(Ref.~\onlinecite{qur2012b}). However, so far anisotropic magnetic
response has only been reported for superconducting samples close
to the antiferromagnetically ordered phase leaving the relevance
of this anisotropy for the superconducting state matter of debate.

 Here we present a polarized
inelastic neutron scattering study on the hole-overdoped compound
Ba$_{0.5}$K$_{0.5}$Fe$_2$As$_2$. Besides their rather high
superconducing transition temperatures the K over-doped samples
appear interesting as they bridge the superconducting state at
optimum doping \cite{rot2008b,urb2010,avc2012} supposed to exhibit
$s\pm$ symmetry with that in KFe$_2$As$_2$ \cite{liu2008} for
which other order parameter symmetries were
proposed.\cite{rei2012} Concomitantly, the magnetic response
changes from commensurate excitations at x=0.33
\cite{wang2013,zhang2013} to longitudinally modulated ones
observed in KFe$_2$As$_2$. \cite{lee2011} So far the intermediate
doping range was only studied by powder neutron scattering
experiments reporting the evolution of the longitudinal
incommensurabilities.\cite{cas2011} Although the doping level in
Ba$_{0.5}$K$_{0.5}$Fe$_2$As$_2$ is located considerably above the
value where antiferromagnetic order fully disappears, we have
found that this compound still exhibits signatures of the ordered
phase: There is sizeable $q_l$ dispersion and most importantly a
well defined spin-space anisotropy develops below the resonance
energy.


Single crystals of Ba$_{0.5}$K$_{0.5}$Fe$_2$As$_2$ were grown by
the self-flux method.\cite{kih2010} The critical temperature of
the single crystals was determined to be 36 K from the temperature
dependence of the zero-field-cooled magnetization studied on
several individual single crystals with a SQUID magnetometer. The
$c$ lattice constant of each single crystal was examined by x-ray
diffraction on both sides of the tabular-shaped samples. We find
only minor variation with values staying between 13.40 and
13.45~\AA indicating a variation of the K content below
$\Delta$x=$\pm$0.028. A total number of 60 single crystals with a
total mass of 1.3 g were co-aligned on a thin Al sample
holder.\newline Polarized inelastic neutron scattering experiments
with longitudinal polarization analysis were carried out at the
thermal-beam spectrometers IN20 and IN22 (ILL, Grenoble). Both
spectrometers were equipped with polarizing Heusler (111) crystals
as monochromator and analyzer. The flipping ratio was determined
on a nuclear reflection to be approximately 17 at IN20 and 13 at
IN22, respectively. All inelastic scans were performed with a
constant $\mathbf{k}_f$ of 2.662~\AA$^{-1}$. The sample was
mounted in the [110]/[001] scattering plane. Longitudinal
polarization analysis was performed using the CRYOPAD device to
guide and orient the neutron polarization with a strictly zero
magnetic field at the sample position in order to avoid errors due
to flux inclusion and field repulsion in the superconducting state
of the sample. We use the common coordinate system in polarized
neutron scattering \cite{pol-neutr} with $x$ pointing along the
scattering vector, $y$ being perpendicular to $x$ within the
scattering plane, and $z$ pointing perpendicular to the scattering
plane. As a general law neutron scattering only senses magnetic
excitations polarized perpendicular to the scattering vector
$\mathbf{Q}$. With the polarization analysis it is possible to
separate nuclear scattering [always a non-spin-flip (NSF) process]
from magnetic scattering and to separate magnetic fluctuations
polarized along different direction in spin space. Magnetic
excitations contribute to the spin-flip (SF) channel only for
magnetic components oscillating perpendicular to the initial
polarization axis $\mathbf{P}_i$. In contrast, magnetic
excitations with components oscillating parallel to $\mathbf{P}_i$
are detected in the NSF channel. For each point in the scan the
three SF channels and the NSF$_x$ channel have been measured, the
latter being a reference for spurious scattering. The respective
magnetic cross sections for the in-plane and out-of-plane
response, $\sigma_z$ and $\sigma_y$, respectively, can be deduced
by simple algebra and by correcting for the finite flipping ratio
as it is explained in detail in Ref.~\onlinecite{qur2012b}.

\begin{figure}
\includegraphics[width=0.44\textwidth]{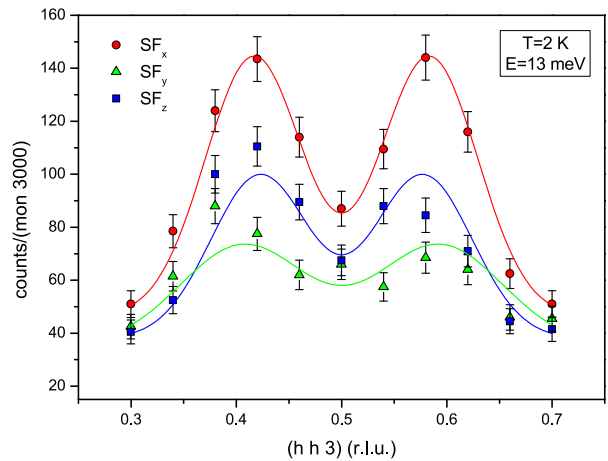}
\caption{\label{fig:qscan} (Color online) Longitudinal
constant-energy scan at $E$=13 meV across $\mathbf{Q}$=(0.5 0.5 3)
(measured at the IN20 spectrometer, monitor 3000 corresponds to
580 s of beam time at $E$=13 meV) proving the incommensurability of the magnetic
resonance peak.}
\end{figure}

\begin{figure}
\includegraphics[width=0.49\textwidth]{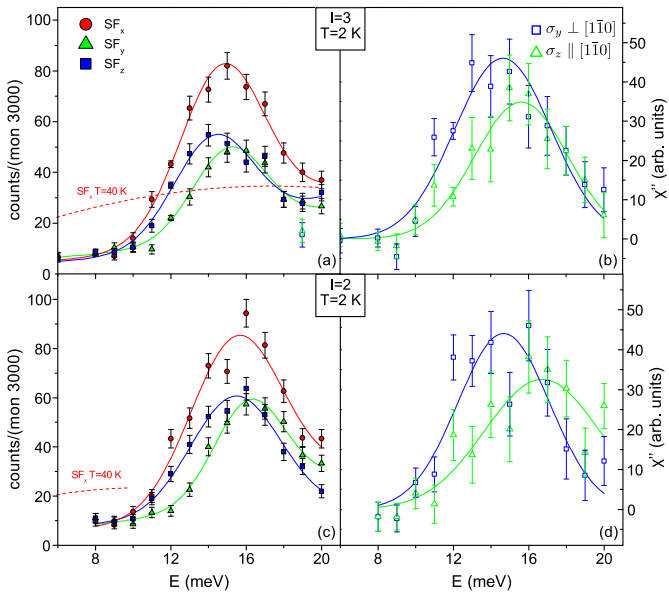}
\caption{\label{fig:escans} (Color online) Constant-$\mathbf{q}$
scans at $\mathbf{Q}$=(0.56 0.56 3) (first row) and
$\mathbf{Q}$=(0.56 0.56 2) (second row) at 2 K within the
superconducting phase (measured at the IN22 spectrometer). (a) and
(c) show the raw intensities of the SF channels together with the
SF$_x$ scattering at $T$=40 K [(red) dashed line]. The fit of a
Gaussian function on a second degree polynomial background to the
raw data is represented by solid lines. In (b) and (d) the
imaginary part of the magnetic susceptibility is depicted after
correcting the raw intensities as described in the text. The solid
lines are Gaussian fits on a zero background (monitor 3000
corresponds to 530 s of beam time at $E$=8 meV and 680 s at $E$=20 meV on IN22, this also applies to
figures 3-5). }
\end{figure}
\begin{figure}
\includegraphics[width=0.44\textwidth]{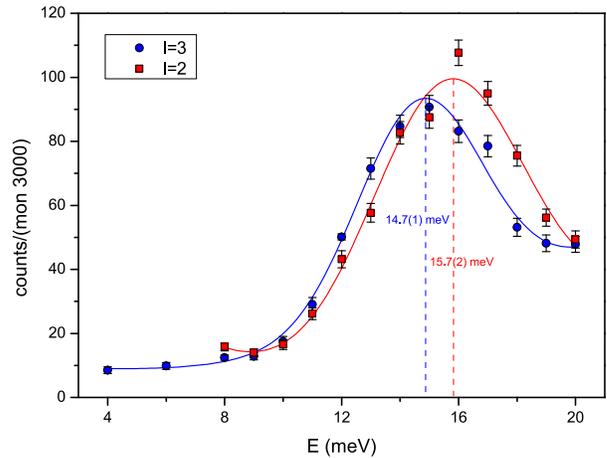}
\caption{\label{fig:disp} (Color online) Total magnetic  intensity
obtained by (SF$_x$+SF$_y$+SF$_z$)/2 indicating a weakly
dispersive magnetic resonance peak. The solid lines are Gaussian
functions on a second degree polynomial background fitted to the
raw data. From the peak positions a bandwidth of 1 meV can be
obtained (errors given in brackets correspond to the preceding
digit). }
\end{figure}

As a first step of the characterization of the magnetic
fluctuations in Ba$_{0.5}$K$_{0.5}$Fe$_2$As$_2$ a longitudinal
constant-energy scan was performed at the IN20 spectrometer in
order to derive the incommensurability of the magnetic signal.
Fig.~\ref{fig:qscan} shows the raw data at an energy transfer of
13 meV for the three SF channels. The longitudinally split peaks
are clearly  visible in the purely magnetic SF channels. We have
derived an incommensurability of 0.083(2) reduced lattice units
for E=13\ meV from a fit of a pair of symmetrical Gaussian
functions to the SF$_x$ data. The incommensurate character of the
scattering and the value of the pitch perfectly agree with the
data obtained on powders \cite{cas2011} and with our own
unpolarized neutron studies  \cite{lee2015}; with the doping the
shape of the Fermi surface sheets and therefore the nesting
conditions change inducing incommensurate
scattering.\cite{cas2011} For the energy scans aiming to detect
the spin anisotropy we chose the average incommensurability
resulting from the unpolarized experiments, i.e.
$\mathbf{Q}$=(0.56 0.56 $q_l$). Fig.~\ref{fig:qscan} shows,
furthermore, significant anisotropy with the scattered intensity
in the SF$_z$ channel being stronger than that in the SF$_y$
channel. In order to reveal the energy dependence of this
anisotropy constant-$\mathbf{Q}$ scans at $\mathbf{Q}$=(0.56 0.56
3) [Fig.~\ref{fig:escans}(a)-(b)] and at $\mathbf{Q}$=(0.56 0.56
2) [Fig.~\ref{fig:escans}(c)-(d)] were carried out at $T$=2 K
(Note that the data in this and all following figures have been
obtained on the IN22 spectrometer, however, the IN20 spectrometer
yields the same results). Figs.~\ref{fig:escans}(a) and (c) show
the raw data in the SF channels together with a polynomial fit to
the SF$_x$ scattering at 40 K.
A clear magnetic signal is observable around 15 meV forming the
broad magnetic resonance peak as a consequence of the opening of
the superconducting gap and the redistribution of spectral weight
indicated by the SF$_x$ scattering at $T>T_C$.
Figs.~\ref{fig:escans}(b) and (d) show the magnetic cross sections
$\sigma_{y,z}$ obtained by subtracting the SF channels following
Ref.~\onlinecite{qur2012b} and by correcting for the finite
flipping ratio, the Bose factor and higher order contaminations at
the monitor, yielding the imaginary part of the magnetic
susceptibility. Similar to previous reports on other FeAs
compounds the excitations corresponding to $\sigma_y$ set in at
lower energy than the transverse in-plane ones corresponding to
$\sigma_z$. Furthermore, the raw data [Figs.~\ref{fig:escans}(a) and (c)] suggest a weak dispersion for both the in-plane and out-of-plane response of the
magnetic resonance peak between the scattering vector with odd
$q_l$=3 and that with even $q_l$=2. In order to quantify the band
width the total magnetic signal (SF$_x$+SF$_y$+SF$_z$)/2 has been
analyzed which is shown in Fig.~\ref{fig:disp}. From the fit of a
Gaussian function on a polynomial background to the total magnetic
signal with $q_l$=3 and $q_l$=2 we have derived a band width of
approximately 1 meV, see also Ref.~\onlinecite{lee2015}\newline

\begin{figure}
\includegraphics[width=0.44\textwidth]{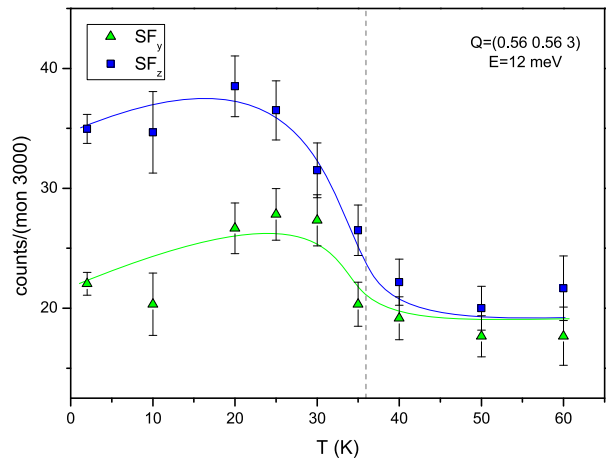}
\caption{\label{fig:tdep} (Color online) Peak intensity of the magnetic scattering
at $\mathbf{Q}$=(0.56 0.56 3) and $E$=12 meV as a function of temperature. A clear
splitting of the in-plane ($SF_y$) and the out-of-plane ($SF_z$) components is visible
at $T_c$ which is indicated by the dashed line. The solid lines are guides to the eye.}
\end{figure}

In order to prove that the spin-space anisotropy is connected to
the superconducting state we have followed the scattered intensity
at $\mathbf{Q}$=(0.56 0.56 3) and an energy transfer of 12 meV,
where according to Fig.~\ref{fig:escans}(a) a pronounced
anisotropy is present, as a function of temperature.
Fig.~\ref{fig:tdep} shows the scattering in the two SF channels.
No difference within the error bars is visible above $T_c$
(indicated by the dashed line) indicating an isotropic system. At
cooling below $T_c$  a significant spin-space anisotropy emerges.
\newline

\begin{figure}
\vskip4mm
\includegraphics[width=0.495\textwidth]{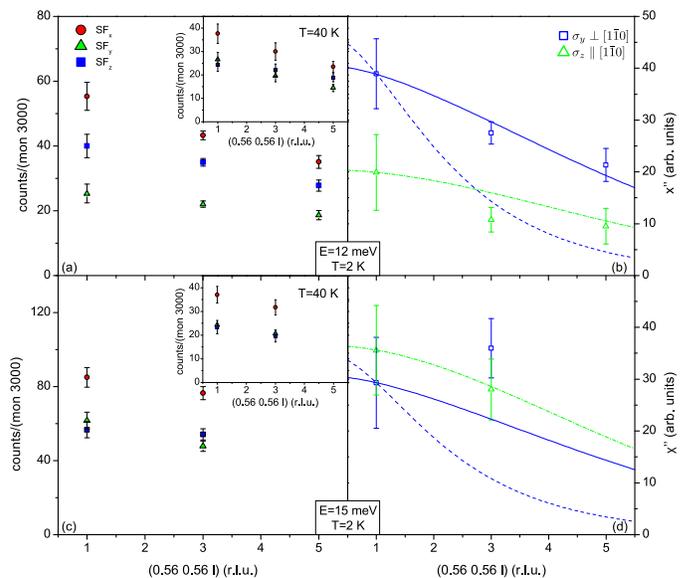}
\caption{\label{fig:ldep} (Color online) $q_l$ dependence of the
magnetic signal at $\mathbf{Q}$=(0.56 0.56 $q_l$) and $T$=2 K for
an energy transfer of 12 meV (first row) and 15 meV (second row).
(a) and (c) show the raw intensities of the SF channels (the insets show the data at $T$=40 K). (b)
and (d) show the magnetic cross sections $\sigma_{y,z}$. The
(green) dash-dotted line in (b) represents the magnetic form
factor normalized to the $\sigma_z$ data point at $q_l$=1, the
(blue) dashed line is the geometrical factor (only containing the
transversal component of the fluctuation) times the magnetic form
factor normalized to the $\sigma_y$ data point at $q_l$=1 and the
(blue) solid line is a fit containing the geometrical factor (both
the longitudinal and the transversal part of the fluctuation)
times the magnetic
 form factor normalized in the same way. In (d) the (blue) solid line shows the fit in (b) adjusted to
 the $\sigma_y$ $q_l$=1 point due to the small extent in $q_l$ range.}
\end{figure}

The polarization analysis directly distinguishes between $\sigma
_z$ always sensing the transversal in-plane magnetic contributions
and $\sigma _y$ sensing the fluctuations parallel to the neutron
scattering plane. Further information can be obtained by varying
the scattering vector ${\bf Q}$=(0.56,0.56,$q_l$): For $q_l$=0
only out-of-plane fluctuations can contribute to $\sigma _y$, but
this out-of-plane contribution decreases with increasing $q_l$
while the in-plane longitudinal contribution increases with the
geometry factor $sin(\alpha)^2$=$\left({\frac{q_l}{\vert Q
\vert}}\right)^2$ (q-components given in absolute units here;
$\alpha$ the angle between [110] and the scattering vector {\bf
Q}). Fig.~\ref{fig:ldep} shows the magnetic signal at
$\mathbf{Q}$=(0.56 0.56 $q_l$) for odd $q_l$ values up to 5.
Fig.~\ref{fig:ldep}(a) clearly reveals the anisotropy which we
have observed at E=12 meV up to $q_l$=5. In contrast,
Fig.~\ref{fig:ldep}(c) and the inset of Fig.~\ref{fig:ldep}(a)
confirm the isotropic fluctuations at an energy transfer of 15 meV
and above $T_c$, respectively. In order to analyze the respective
intensities quantitatively we have converted the raw data to the
imaginary part of the magnetic susceptibility [see
Fig.~\ref{fig:ldep}(b)]. Let us first consider the magnetic cross
section $\sigma_z$ measured in the SF$_y$ channel, which is not
subject to the geometrical factor with varying $q_l$ value.
Therefore, an increase in $q_l$ will affect the intensities by the
magnetic form factor $f(\mathbf{Q})$ only. In fact, $\sigma_z$ is
reduced proportional to $f(\mathbf{Q})^2$ which is shown by the
(green) dash-dotted line normalized to $\sigma_z$($q_l$=1). As
mentioned above, the out-of-plane contributions in the SF$_z$
channel are reduced by an increasing $q_l$ value, therefore, the
(blue) dashed line represents $f(\mathbf{Q})^2\cos^2\alpha$, with
$\alpha$ the angle between Q and the [110] direction, normalized
to the data point $\sigma_y$($q_l$=1). It becomes evident that a
purely out-of-plane fluctuation is reduced more severely, but our
data still shows $\sigma_y>\sigma_z$ at $q_l$=5. Such a behavior
can only be explained by a sizeable in-plane longitudinal
contribution of magnetic fluctuations. We have fitted the 12\ meV
data with $f(\mathbf{Q})^2[n_{t}\cos^2\alpha+(1-n_t)\sin^2\alpha]$
normalized to the intensity at $l=1$, where $n_t$ defines the
fraction of the $\sigma_y$ channel intensity which stems from the
out-of-plane fluctuation. The fit reveals a value of $n_t$=0.53(7)
stating that the in-plane longitudinal fluctuation is of
comparable strength as the out-of-plane fluctuation. On the
contrary, at this energy of 12\ meV the transversal in-plane
fluctuations are reduced by roughly a factor two. Transversal
in-plane magnetic fluctuations seem to appear at higher energy.
Therefore, the character of the magnetic correlations in
Ba$_{0.5}$K$_{0.5}$Fe$_2$As$_2$ can be considered as an easy-plane
type (two soft and one hard axes). Spin orbit coupling is clearly
a relevant parameter for a quantitative understanding of magnetic
excitations in superconducting FeAs-based compounds even when they
are not close to the antiferromagnetic phase.
Fig.~\ref{fig:ldep}(d) shows that there seems to be no anisotropy
in the spin excitations at 15meV, whose $q_l$ dependence is well
described by the magnetic form factor.

In conclusion we have presented inelastic neutron scattering data
with longitudinal polarization analysis revealing  anisotropic and
dispersive magnetic excitations in superconducting
Ba$_{0.5}$K$_{0.5}$Fe$_2$As$_2$. The investigated sample is in the
overdoped region of the phase diagram\cite{avc2012} and therefore
far away from the magnetically ordered phase. Nevertheless the
anisotropic magnetic character of the parent
compound\cite{qur2012b} persists as we find $c$ polarized spin
fluctuations at lower energies than the transverse in-plane
polarized modes. Spin space anisotropy induced by spin-orbit
coupling is thus relevant in a broad part of the superconducting
phase and not only near the existence of static antiferromagnetic
order. By analyzing the $q_l$ dependence of magnetic excitations
we can furthermore identify pronounced in-plane anisotropy.
Magnetic correlations in Ba$_{0.5}$K$_{0.5}$Fe$_2$As$_2$ exhibit
an easy-plane or hard-axis character, like it is the case for
electron doping.\cite{luo2013,ste2013,wasser} This magnetic
anisotropy must arise from spin orbit coupling and a peculiar
orbital arrangement. Furthermore, in light of the universality of
the dispersive spin-resonance mode among the FeAs
superconductors,\cite{lee2013} Ba$_{0.5}$K$_{0.5}$Fe$_2$As$_2$
exhibits significant $q_l$ dispersion of the magnetic resonance.

\begin{acknowledgments}
This study was supported by a Grant-in-Aid for Scientific Research
B (No. 24340090) from the Japan Society for the Promotion of
Science and by the Deutsche Forschungsgemeinschaft through the
Priority Programme SPP1458 (Grant No. BR2211/1-1).

\end{acknowledgments}


\end{document}